\documentclass[a4paper,11pt,preprint,3p]{elsarticle}
\usepackage[utf8]{inputenc}
\usepackage{graphicx}
\usepackage{float}
\usepackage{adjustbox}
\usepackage{lineno}
\usepackage{setspace}
\usepackage{xcolor}
\usepackage{url}
\title{Reconstruction of Fast Neutron Direction in\\Segmented Organic Detectors using Deep Learning}
\author{Jun Woo Bae, Tingshiuan C. Wu and Igor Jovanovic}
\affiliation{organization={Department of Nuclear Engineering and Radiological Sciences,\\University of Michigan},
addressline={2355 Bonisteel Blvd},
city={Ann Arbor},
postcode={MI 48109},
country={United States}}

\date{August 2022}

\begin{document}

\begin{abstract}
    A method for reconstructing the direction of a fast neutron source using a segmented organic scintillator-based detector and deep learning model is proposed and analyzed. The model is based on recurrent neural network, which can be trained by a sequence of data obtained from an event recorded in the detector and suitably pre-processed. The performance of deep learning-based model is compared with the conventional double-scatter detection algorithm in reconstructing the direction of a fast neutron source. With the deep learning model, the uncertainty in source direction of 0.301~rad is achieved with 100 neutron detection events in a segmented cubic organic scintillator detector with a side length of 46~mm. To reconstruct the source direction with the same angular resolution as the double-scatter algorithm, the deep learning method requires 75\% fewer events. Application of this method could augment the operation of segmented detectors operated in the neutron scatter camera configuration for applications such as special nuclear material detection.

\end{abstract}
\maketitle
\section{Introduction}

Directional detection of fast neutrons is a technique that has seen significant attention in the search for special nuclear material (SNM) for nuclear security and safeguards applications. For directional detection of fast neutrons, a neutron scattering camera technique that employs segmented detectors in two independent detection volumes~\cite{direc_vanier2007, scattercamera_Mascarenhas2009, doublescatter_Steinberger2020}, a single-volume detector~\cite{doubleScatter_braverman2018, singlevolume_Manfredi2020, doubleScatter_Galindo_Tellez_2021}, or a time-projection chamber~\cite{direc_BOWDEN20101, direc_Igor2009} has been devised. The method of double-scatter backprojection relies on the kinematics by recording the energy and position of particles before and after scattering~\cite{scatter_STEVANATO2011}.
In an organic scintillator, even if the light output-energy relationship is precisely calibrated for recoil protons, the deposited energy does not always directly relate to the recorded scintillation light output because of the sequence of multiple scatters through which the energy can be deposited combined with detector nonlinearity, along with the relatively poor energy resolution~\cite{scint_KASCHUCK2002}. For this reason, when a particle scatters in two separate detection planes, a method of calculating the speed, that is, energy of a neutron,  uses the known distance and measured time-of-flight (ToF) between those two planes~\cite{doubleScatter_Zhang2016}. In addition to the fact that the double-scatter events are relatively rare because scatters must occur in both detection planes for the same incident neutron, the reconstruction accuracy can be limited by the achievable time resolution. In order to reconstruct the source direction, many detector interactions are required, as most events typically produce only single scatters in a compact detector.
This motivates the development of a directional detection method that could employ the information provided by a greater fraction of all neutron interactions with the detector and not only those that result in double scatter. This paper explores the application of machine learning to take advantage of the information provided by the entire range of neutron interactions and topologies recorded in a segmented detector to reconstruct the incident neutron direction.

Machine learning is now used to solve a variety of problems in physical sciences~\cite{machinelearning_2019}. In particle physics, where massive amounts of data need to be interpreted, the machine learning has already been recognized as a powerful tool and is increasingly adopted~\cite{machinelearning_Schwartz2021}. Beyond traditional machine learning models that utilize statistical correlation of parameters, deep learning models that use neural networks have become more versatile and can be trained using high-dimensional data or time-dependent sequential data~\cite{deeplearning_Shengdong2018}. In the field of ionizing radiation measurement and high-energy physics, deep learning has been used for nuclide identification~\cite{deeplearning_GOMEZFERNANDEZ2021} and particle classification~\cite{deeplearning_carminati2017}. In these applications, artificial neural networks are used for spectroscopic analysis, to detect rare events, and to examine the correspondence between the experimental and simulated data by utilizing generative adversarial networks.

In this study, we present the design and performance of a deep learning model in reconstructing the direction of an isolated source of fast neutrons. The model uses a series of detector signals that comprise events within a detector volume that provides position sensitivity through detector segmentation. The method arranges event information in the form of a sequence data and uses this data to train a recurrent neural network-based deep learning model. The training and test data are generated by simulation, and the performance of the method is compared to the double-scatter backprojection method that uses the same data.
    
\section{Materials and Methods} \label{Materials and Methods}

\subsection{Simulation Overview} \label{Simulation Overview}

A single or a series of interactions in a detector induced by a single incident neutron constitutes an event. Each event carries a certain degree of information related to the direction of the incident fast neutron. In order to reconstruct the angle of incidence of fast neutrons, a series of signals can be used that originate from recoil protons produced by neutron scattering in a hydrogen-rich material such as an organic scintillator. The energy transfer to recoil proton is modeled as a Monte Carlo particle transport simulation and converted to light output such that it is proportional to $E^{3/2}$~\cite{knoll2010radiation}. Detector segmentation in 3 dimensions is used to achieve position resolution; as a result, the following parameters can be associated with the scintillation detector signal: light output, time, and position (which is discretized to a detector segment).
The detected signal is stored as sequential data according to the chronological order in which the energy deposition occurs. The sequence data have a length of 16 and contain 5 values: light output, time, and a discrete position of the detector segment ($i, j, k)$.
There are several cases of interactions that produce ambiguous input as sequence data  but occur relatively infrequently. In the first case, multiple interactions occur in the same detector segment. The duration of an event was set to 600~ns, which is of the order of a typical digitized waveform for an organic scintillator combined with a silicon photomultiplier. If interactions take place within this interval, it is assumed that they cannot be distinguished in real-time data processing. Therefore, the light output produced by the multiple interactions in this case are added. The interaction beyond this time interval is rejected and not stored in the sequence data. In the second case, interactions are recorded to occur at the same time because of limited time resolution. In this case, the interaction occurring closer to the face of the detector is stored as the first in a sequence. 
\begin{figure}[H]
    \centering
    \includegraphics[scale = 0.6]{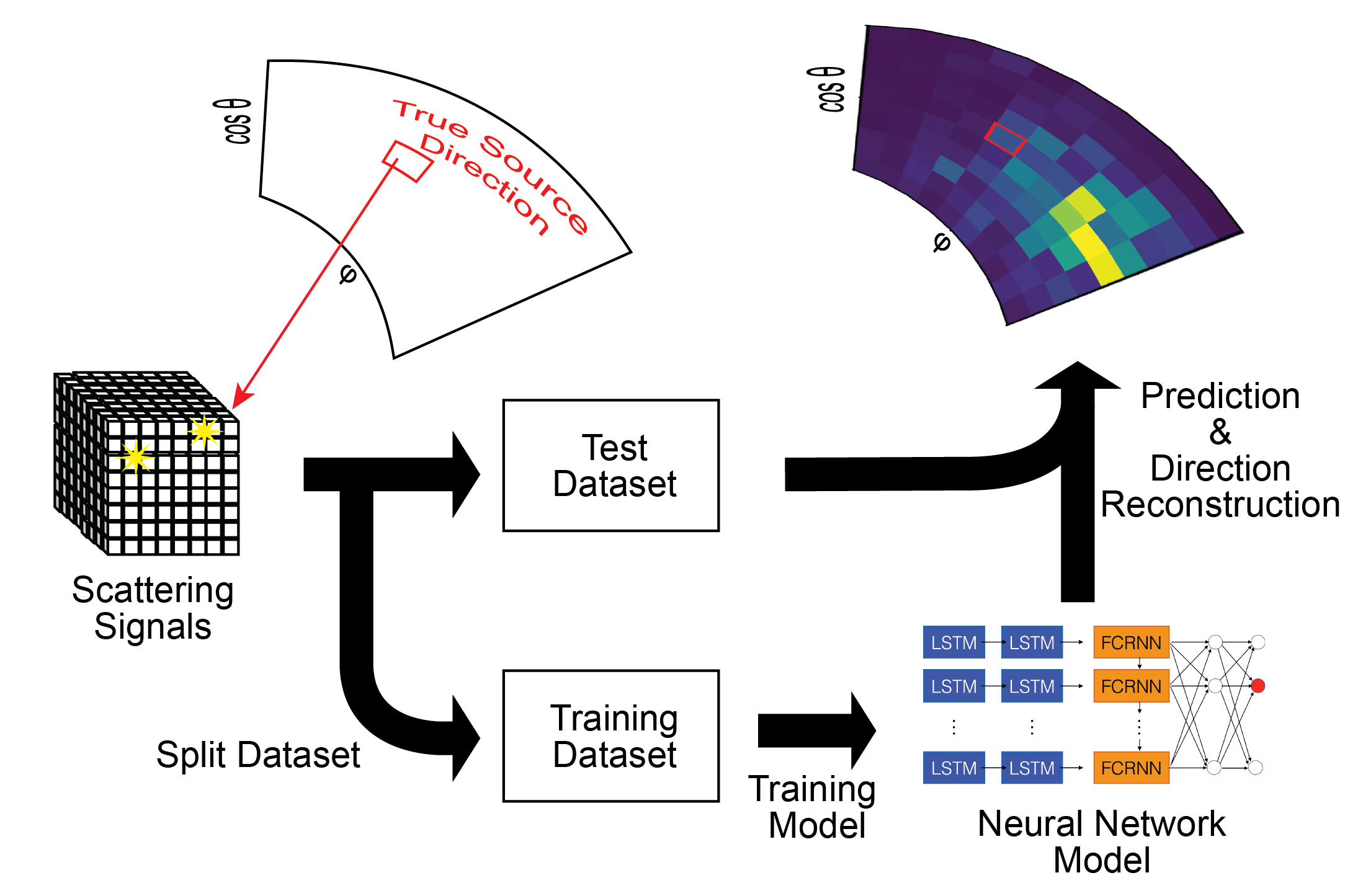}
    \caption{Illustration of the  process of fast neutron direction reconstruction using deep learning. The proton recoil represents a detector signal. The signal is stored into a dataset, which is split into test and training datasets. The model is pre-trained by the training dataset, and it is validated with the test dataset.}
    \label{fig:1}
\end{figure}

Figure~\ref{fig:1} shows the reconstruction process of fast neutron using deep learning. A dataset is prepared using a standard radiation particle transport simulation code, the GEANT4 (GEometry ANd Tracking) simulation toolkit~\cite{Geant4_AGOSTINELLI2003}. We employed a dataset with $1.3 \times 10^5$ detection events resulting from $10^6$ incident fast neutrons; 80$\%$ of dataset is used for training of a deep learning model, while the remainder is used as a test dataset. The full solid angle seen by the detector is divided into 10 equal polar and 10 azimuthal segments, such that each angular segment subtends the same azimuthal angle ($\phi$) and the same cosine of polar angle ($\cos{\theta}$). The performance of the trained model is evaluated with randomly sampled events from the test dataset.

\subsection{Detector Geometry and Source Definition} \label{Detector Geometry and Source Definition}

\begin{figure}[H]
    \centering
    \includegraphics[scale = 0.4]{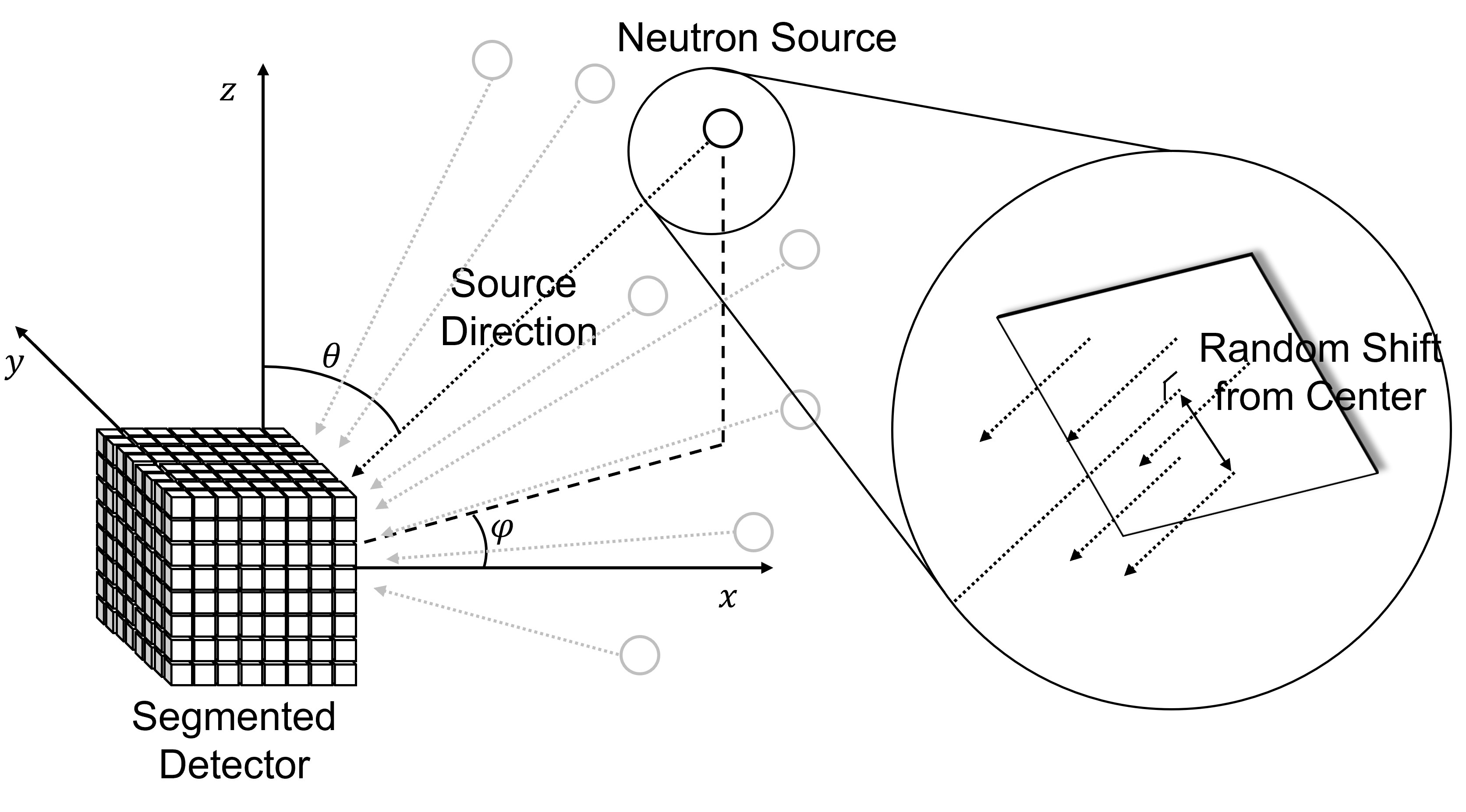}
    \caption{Detector and source geometry used for the simulation. The direction of a neutron is randomly sampled such that it is directed towards the center of the detector, and its position is shifted parallel to this direction to emulate \textit{far~field detection}.}
    \label{fig:2}
\end{figure}

Figure~\ref{fig:2} shows the detector and source geometry used for the simulation. The simulated detector is a 8 $\times$ 8 $\times$ 8 3-D segmented detector plastic scintillator with a density of 1.023~g/cm\textsuperscript{3} and H:C ratio of 1.107 ; this data is obtained from the GEANT4/NIST material database~\cite{plastics_SC_PVT}. The detector material and geometry are based on that used in Ref.~\cite{sandd_sutanto2021}, and the physical segmentation is extended from 2 to 3 dimensions. Each cell of detector is a cube with a side length of 5.4~mm, and the spacing among cells is 0.4~mm; the entire detector assembly is a cube with side length of 46~mm. For this study, a simple monoenergetic 2.45-MeV neutron spectrum is chosen. The source is positioned as shown in Figure~\ref{fig:2}. The initial direction of the neutrons is randomly sampled,
such that the azimuthal and polar angles follow a uniform distribution from $\phi \in [-\pi/2,\pi/2]$ and $\cos{\theta} \in [-1,1]$ First, the source direction is set towards the center of the detector. The source position is then shifted parallel to the set source direction, which avoids directing the source only to the center of the detector at a specific angle. Otherwise, the neutrons emitted at a specific angle would would consistently interact with the same detector segment, leading to a bias between the source direction and the position of interaction in the detector. The range of shift is $\pm\ \sqrt{3}/2\cdot46$~mm such that the entire field-of-view of the cubic detector is covered. This corresponds to the condition usually referred to as \textit{far-field detection}.

\subsection{Deep Learning Model and Training}  \label{Deep learning model}

 Figure~\ref{fig:2} shows the detector and source geometry used for the simulation. The neutron incident direction is discretized such that directional reconstruction can be treated as a classification problem. Earlier studies that attempt to reconstruct the neutron direction using the backprojection algorithm show a typical angular resolution of $\sim$0.14~rad \cite{Direc_Fu2020, angularacc_miniTimeCube2019, sandd_Li2022}, which corresponds to 22 bins over a $1-\pi$ angular range. However, this makes the number of events assigned to each bin small, increasing the computational cost. For this reason, discretization is performed into 10 bins for $\cos(\theta)$ and 10 bins for $\phi$, so that there is a total of 100 possible incident directions. In the test dataset, the source direction is reconstructed by classifying it into one of the discrete angular segments.

From the dataset, we find that a 2.45-MeV neutron had an average of $1.30 \pm 0.58$ and a maximum of 7 interactions with the simulated detector. The length of the sequence frame was therefore set to 16 to ensure a full set of interactions can be stored while limiting the computational cost. Since the input data contain 5 parameters, as described in Section~\ref{Simulation Overview}, the dimension of input data is 16 $\times$ 5. 

\begin{figure}[H]
    \centering
    \includegraphics[scale = 0.3]{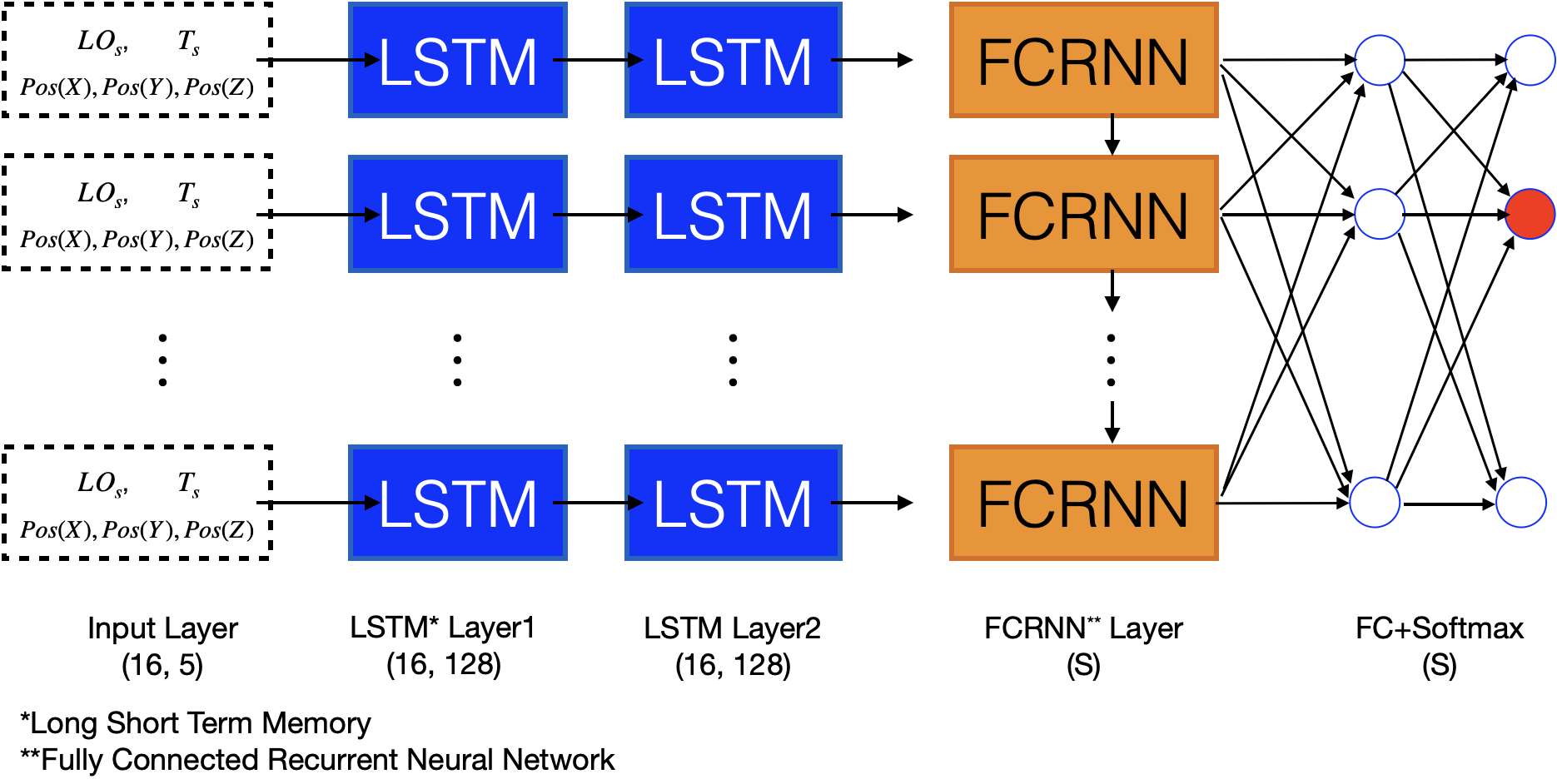}
    \caption{Design schematic of the RNN-based neural network model. The model takes sequence data with a length of 16. The model consists of two LSTM layers, one FCRNN, and fully connected layers.}
    \label{fig:3}
\end{figure}

The recurrent neural network (RNN) is a type of model that can receive a sequential input and learn the dependencies between values in a sequence. This characteristic makes it suitable for capturing the relevant physics contained in the correlations between individual interactions of the interaction sequence comprising an event. Figure~\ref{fig:3} shows a configuration of the model designed for this study. The input layer is followed by two long short-term memory (LSTM) layers and a fully connected recurrent neural network (FCRNN) layer~\cite{RNN_Manaswi2018}. The fully connected layer is followed by a series of RNNs. The number of output nodes of the FCRNN layer is identical to the total number of discrete angular segments (S).

The model calculates the normalized probability of classification in the corresponding angular segment, such that the sum of the probabilities over all segments is unity. The source direction $(\phi, \cos \theta)$ is reconstructed by selecting the maximum probable angular segment.
%
\begin{equation}
\phi_{r}= \phi_i,\;\;  \theta_{r} = \cos^{-1} \left( \cos \theta_i \right),\;\; 
  \{i | i \in  P^{(\phi, \theta)}(i) = max (P^{(\phi, \theta)}(i))\}.
\label{eq1}
\end{equation}
%
%
Here, $P^{(\phi, \theta)}(i)$ is the probability of reconstructing the $i$-th direction segment, $\phi_r$ and $\theta_r$ are the reconstructed azimuthal and polar angles, and $\phi_i$ and $\cos \theta_i$ are the azimuthal angle and cosine polar angle that correspond to the $i$-th angular segment.
The angular accuracy ($\psi$) is defined as the angle between two unit vectors representing the true source direction ($\vec{v}_{t}$) and the reconstructed source direction ($\vec{v}_{r}$):
\begin{equation}
\psi = \cos^{-1} (1-d^2/2), 
\label{eq2}
\end{equation}
\begin{equation}
    d^2 = |\vec{v}_{t}-\vec{v} _{r}|^2. 
    \label{eq3}
\end{equation}

\subsection{Event Updating and Comparison with the Double-Scatter Method}
The deep learning-based method is compared with the double-scatter backprojection method that uses the ToF and distance between scattering events~\cite{doubleScatter_Galindo_Tellez_2021}. The same detector segmented design and simulated event data are used for the comparison. The position of an interaction is assumed to be in the center of a scintillator where it occurs. Bayesian inference is used to update the probability:
\begin{equation}
P^{(n+1)}(i) =  \frac{P^{(*)}(i) \cdot P^{(n)} (i) } {\sum_i^S (P^{(*)} (i) \cdot P^{(n)} (i)) },
\label{eq4}
\end{equation}
where $P^{(n+1)}(i)$, $P_{i}^{(n)}(i)$ and $P_i^{(*)}$ represent probabilities of i-th angular segment for the $(n+1)$-th, $n$-th event update, and the probability of reconstructing the $i$-th angular segment after the $n$-th event.
 In the double-scatter method, the angular cones are calculated as`\cite{doubleScatter_Zhang2016}
\begin{equation}
    \theta_{DS} = \tan^{-1} \sqrt{E_p/E_n},
    \label{eq5}
\end{equation}
\begin{equation}
    E_n = \frac{m}{2} \left( \frac{ \sqrt{\Delta x^2 + \Delta y^2 + \Delta z^2} \cdot l}{\tau} \right)^2.
    \label{eq6}
\end{equation}
Here, $\theta_DS$ is the angle between incident neutron and scattered neutron that defines the angular cone, $E_p$ and $E_n$ are energies of recoil proton and scattered neutron, $m$ is neutron mass, $\Delta x$, $\Delta y$, $\Delta z$ are number of segments between two scattered detector cell in direction of $x$, $y$, and $z$, $l$ is a side length of a detector cell, and $\tau$ is the time of flight between the detector cells in which double scatter takes place. The spatial distance between the two scatter signals is calculated as described in Eq.~(\ref{eq6}), assuming that the signal is generated at the center of the detector cell. The discrete spatial information leads to over- or under-estimation of the energy of scattered neutron as well as error of the vector between two scatter signals. This is an intrinsic limitation in the application of the double scatter method in a detector that has a segmented volume.

 In the RNN method, each event is used to update the reconstructed source direction, whereas only double-scatter events lead to updating the source direction in the case of double-scatter backprojection method. Event-time smearing corresponds to the time resolution of 0, 0.2, and 0.5~ns and is applied and light output signals are cut off with threshold of 0, 0.1, 0.2 MeVee to make the simulation more realistic.

\section{Results and Discussion}

\begin{figure}[H]
        \centering
        \includegraphics[scale = 0.6]{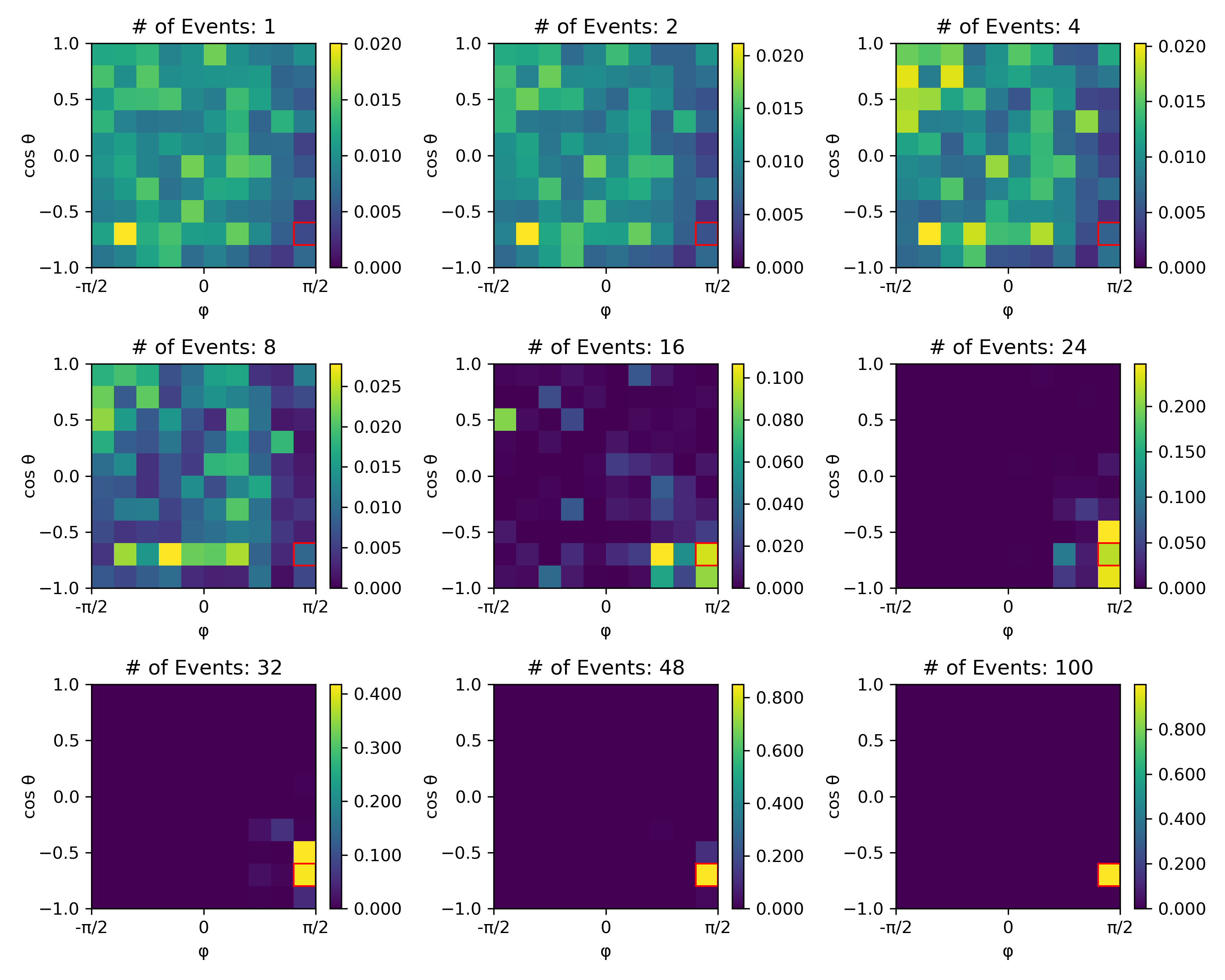}
        \caption{Reconstructed direction of the source neutron using the deep learning model for different numbers of events. The red box indicates the angular segment corresponding to the true direction.}
        \label{fig:4}
\end{figure}

Figure~\ref{fig:4} shows an example of reconstructed source direction using the deep learning model. The probability for each angular segment follows a set color scale, while the red box represents the true source direction that corresponds to $-0.6 < \cos \theta < -0.8$ and $1.26 < \phi < 1.57$~rad, respectively. As the number of events increases from 1 to 48, the maximum probability increases from 0.02 to 0.8. The probability distribution is relatively broad for few events and shrinks with accumulated statistics.

\begin{figure}[H]
        \centering
        \includegraphics[scale = 0.8]{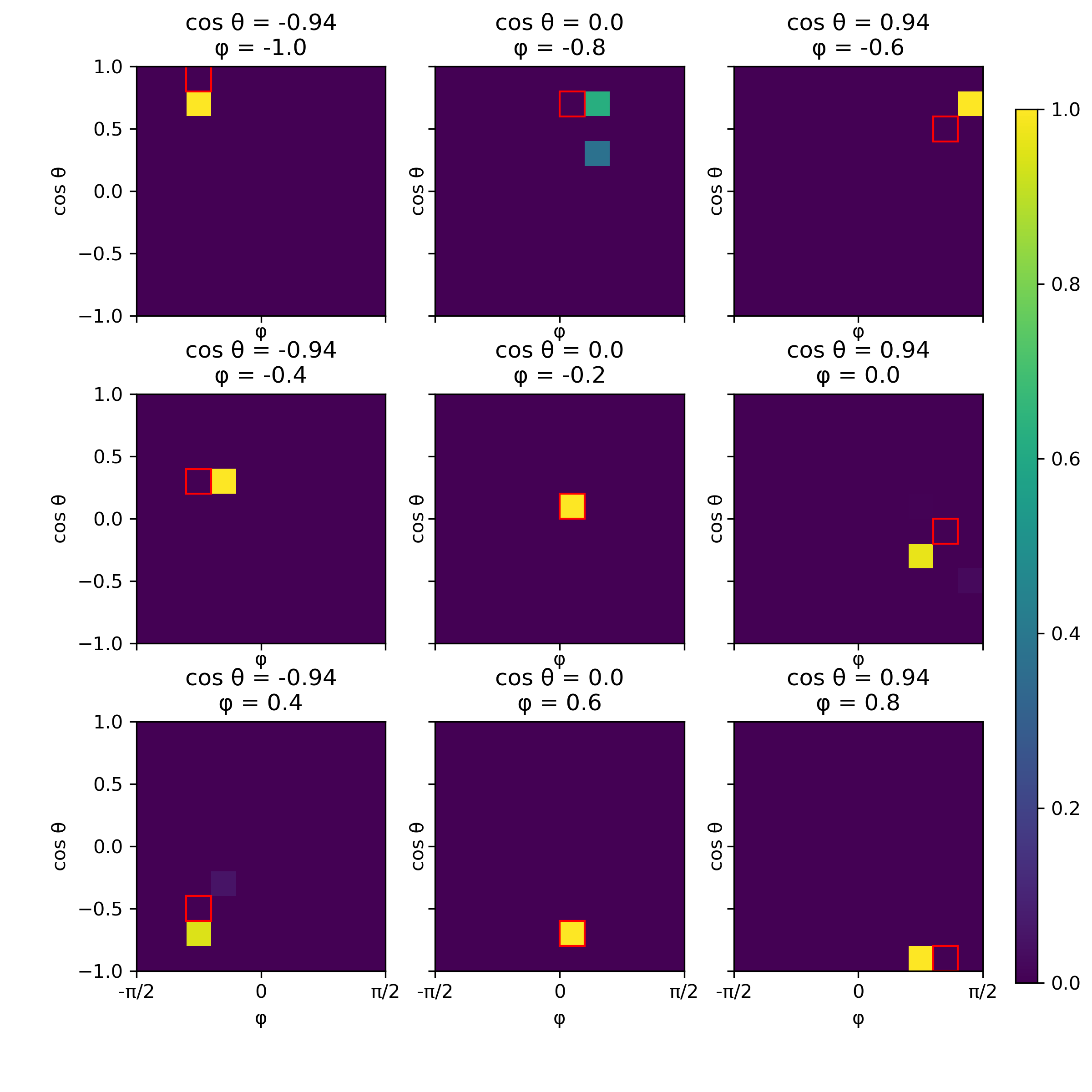}
        \caption{Reconstructed direction using deep learning model with different source directions based on 100 events. The red box and the angular coordinates provided at the top of each panel correspond to the true source direction.}
        \label{fig:5}
\end{figure}

Figure~\ref{fig:5} demonstrates the performance of the RNN model for several incident neutron directions based on 100 events from the test dataset. The red box in each figure represents the true direction, and the corresponding $\cos \theta$ and $\phi$ are presented at the top of each figure. The reconstructed direction matches the true direction. 

\begin{figure}[H]
    \centering
    \includegraphics[scale = 0.8]{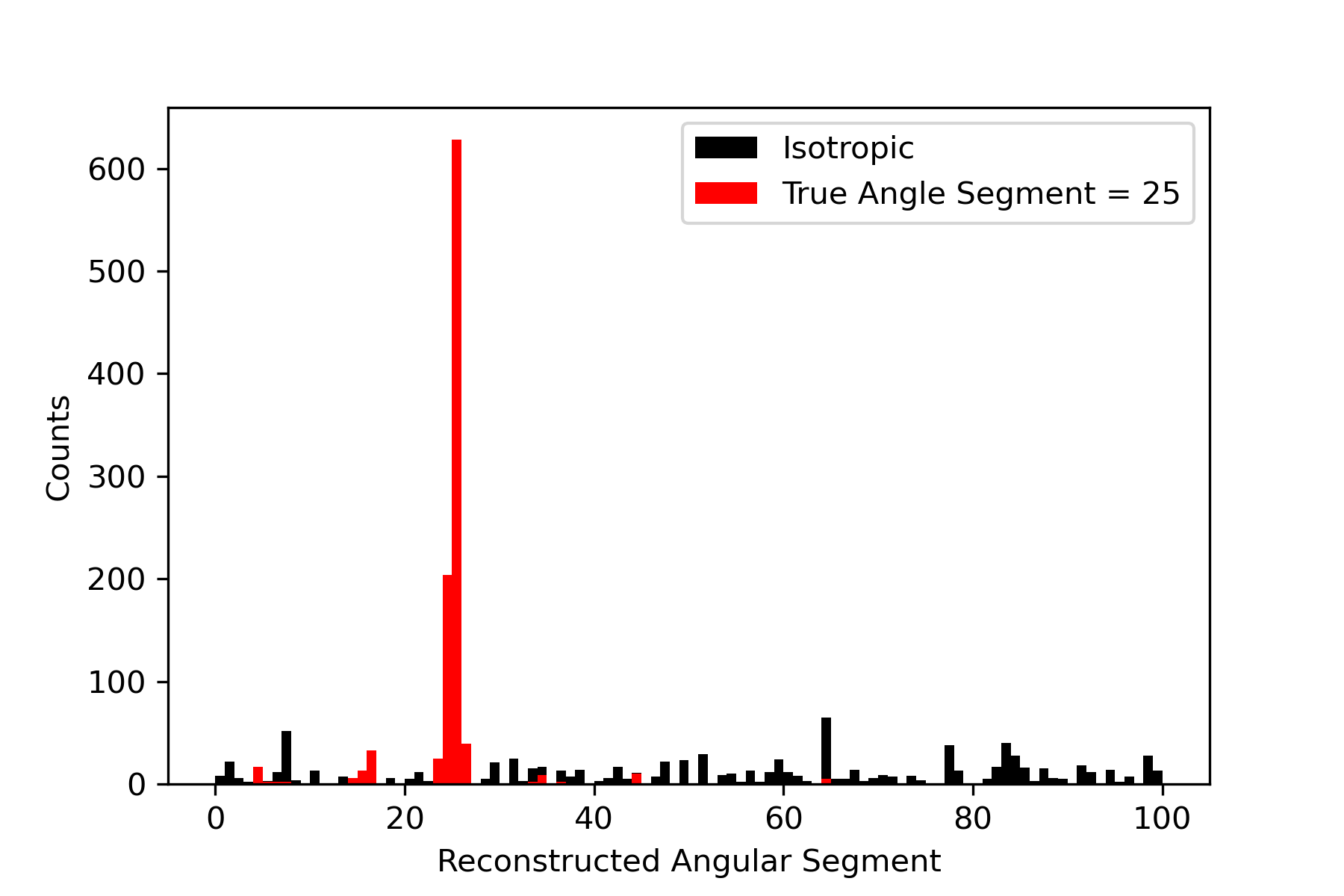}
    \caption{Distribution of reconstructed direction for 1000  trials. The source angle is randomly chosen to emulate background (black), while it is fixed at a single direction for the signal (segment~25 -- red). Each trial uses 100 events.}
    \label{fig:6}
\end{figure}

Figure~\ref{fig:6} shows the distribution of reconstructed directions for two cases: (1)~isotropic case, where the source has no preferred direction (background), and (2)~a source placed at a single direction. In the case of an isotropic source, there is no clear preference for reconstructed direction. In contrast, the reconstruction is consistent in the case of the source emitting from a single direction (segment~25). Small peaks are also shown at segments 15 and 35, 
which are adjacent to the polar angle of the true segment. The Gaussian fit of the peak shows a mean of 24.8 and a full-width half maximum of 1.16. This demonstrates that the trained deep learning model can clearly distinguish between a point source and an isotropic source (background).

\begin{figure}[H]
        \centering
        \includegraphics[scale = 0.8]{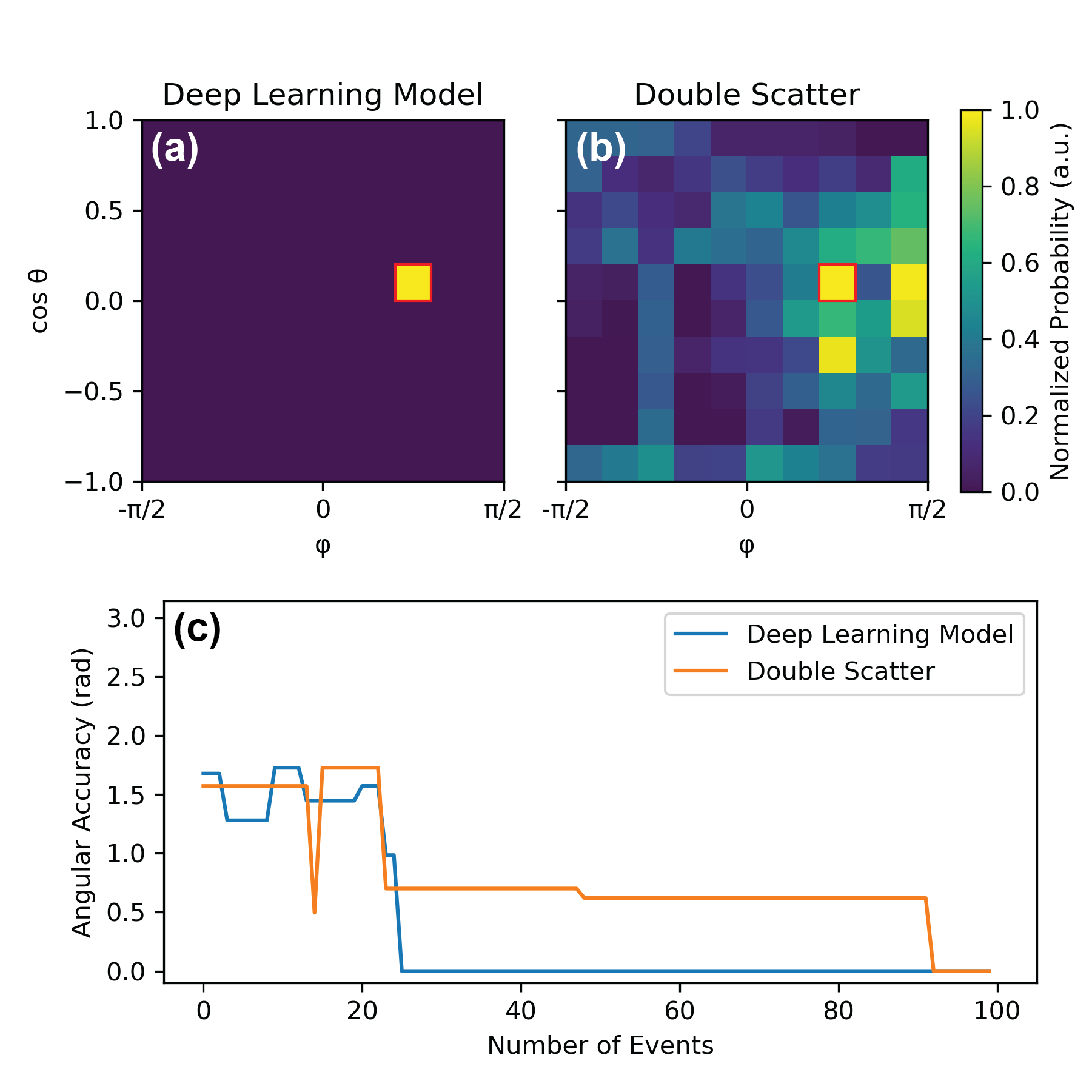}
        \caption{Reconstructed source direction (a)~employing the deep learning model and (b)~the double scatter backprojection method. The red box represents the true direction. (c)~Evolution of angular accuracy for different numbers of events. Input data were randomly sampled from all events in the test dataset.}
        \label{fig:7}
\end{figure}

Figure~\ref{fig:7} shows the comparison of reconstructed source direction using the deep learning model and the double-scatter method, including the evolution of angular accuracy with event statistics. 
In this example, the true source direction of $0 < \cos{\theta} < 0.2$ and  $0.628 < \phi < 0.942$~rad, which is displayed as the red box.
    The angular accuracy in Figure~\ref{fig:7} (c) changes relatively frequently for the deep learning model because it is updated with every event. In contrast, the double-scatter backprojection method leads to step-wise accuracy updates because only a subset of the detected events include double scatters. In the method using the deep learning model, all events can be updated without waiting for the double scatter event, therefore the detection efficiency can be improved, which can improve the sensitivity when used for the detection of special nuclear materials or radioactive isotopes.
It can also be seen that the change in angular accuracy occurs at the same time because both methods update the same event at the same time. When there is a double-scatter event, both the accuracies of the double-scatter method and the deep learning model experience a large change. While a single-scattering event makes a small, incremental change to the reconstructed angle in the deep learning model, a double-scatter event has a much greater impact.

\begin{figure}[H]
    \centering
\includegraphics[scale = 0.8]{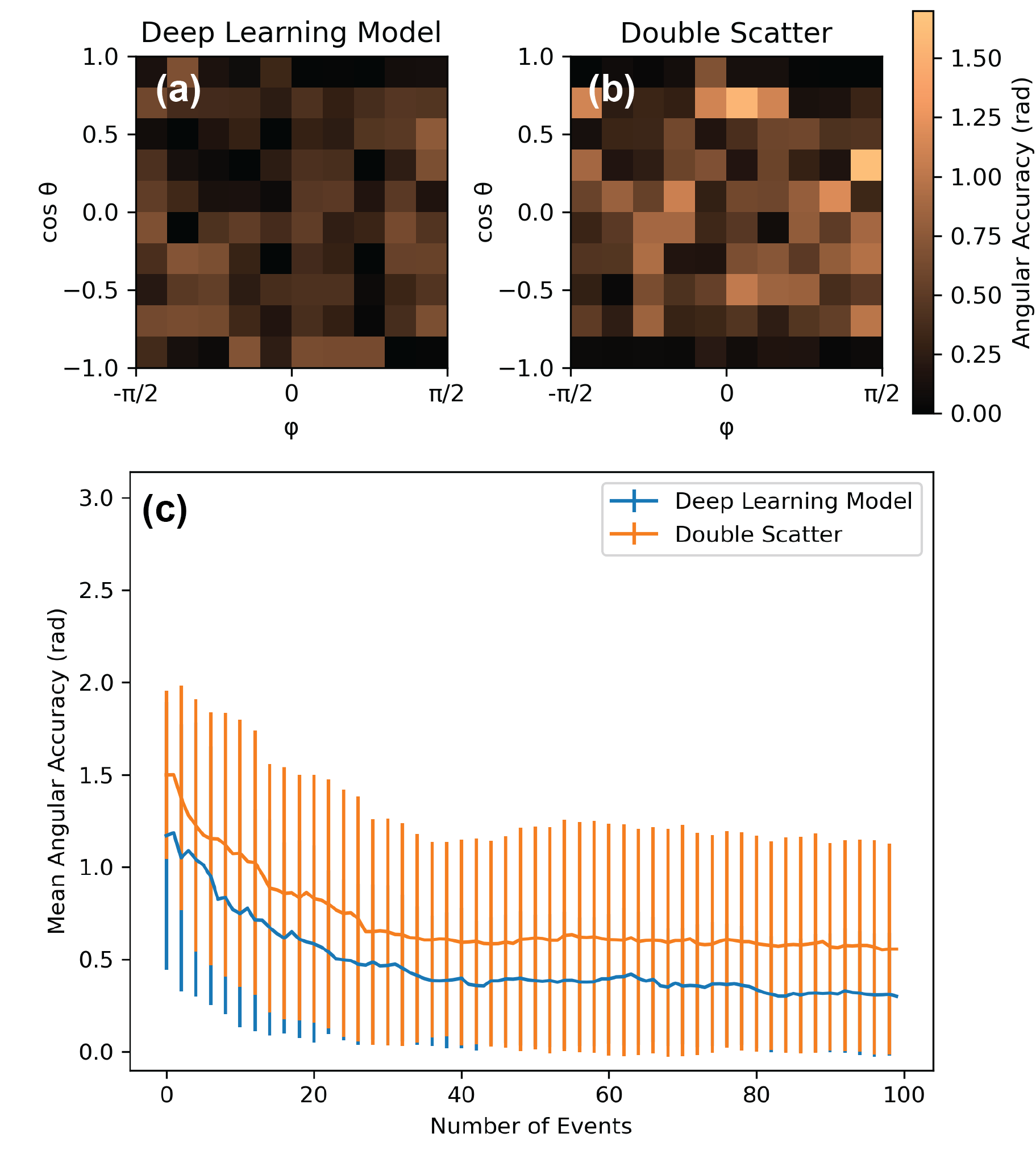}
    \caption{Mean angular accuracy over all angular segments when using (a)~deep learning model and (b)~double-scatter backprojection method. (c)~The evolution of mean angular accuracy is averaged over all angular segments for the deep learning model and the double-scatter backprojection method.}
    \label{fig:8}
\end{figure}

Figure~\ref{fig:8} compares the angular accuracy obtained from the deep learning model and the double-scatter backprojection method. Figure~\ref{fig:8} and (b) shows a distribution of angular accuracy for 100 events for the deep learning model and the double-scatter backprojection algorithm, respectively. The deep learning model shows a better (less dispersed) angular accuracy. In Figure \ref{fig:8} (c), the angular accuracy is averaged over all angular segments. The angular accuracy is different for different angular segments, which is why the average was calculated to evaluate the overall trend for each method. The error bar represents one standard error among angular segments. The mean angular accuracies for the deep learning model and double-scatter backprojection method for 100 events are 0.301 $\pm$ 0.327 and 0.557 $\pm$ 0.571~rad, respectively. The deep learning model reaches the accuracy of the double-scatter backprojection method with $\sim$20 events.  

The miniTimeCube-based neutron scatter camera~\cite{angularacc_miniTimeCube2019} reports the angular difference of ($\theta$, $\phi$)(0.1 and 0.13)~rad with 10,093 neutron events by averaging direction vectors of first two recoils. The iSANDD neutron scatter camera~\cite{sandd_Li2022} reports $\cos \theta=0.99$, which corresponds to the angular difference of 0.14~rad with 5$\times$10\textsuperscript{6} 2.45~MeV neutrons by reconstructing angular cone using the double-scatter backprojection method.   
Because the number of events is different, quantitative comparison is not possible, but the results show that the use of the deep learning model can lead to good accuracy with relatively few ($\sim$100) events.

\begin{figure}[H]
    \centering
    \includegraphics[scale = 0.5]{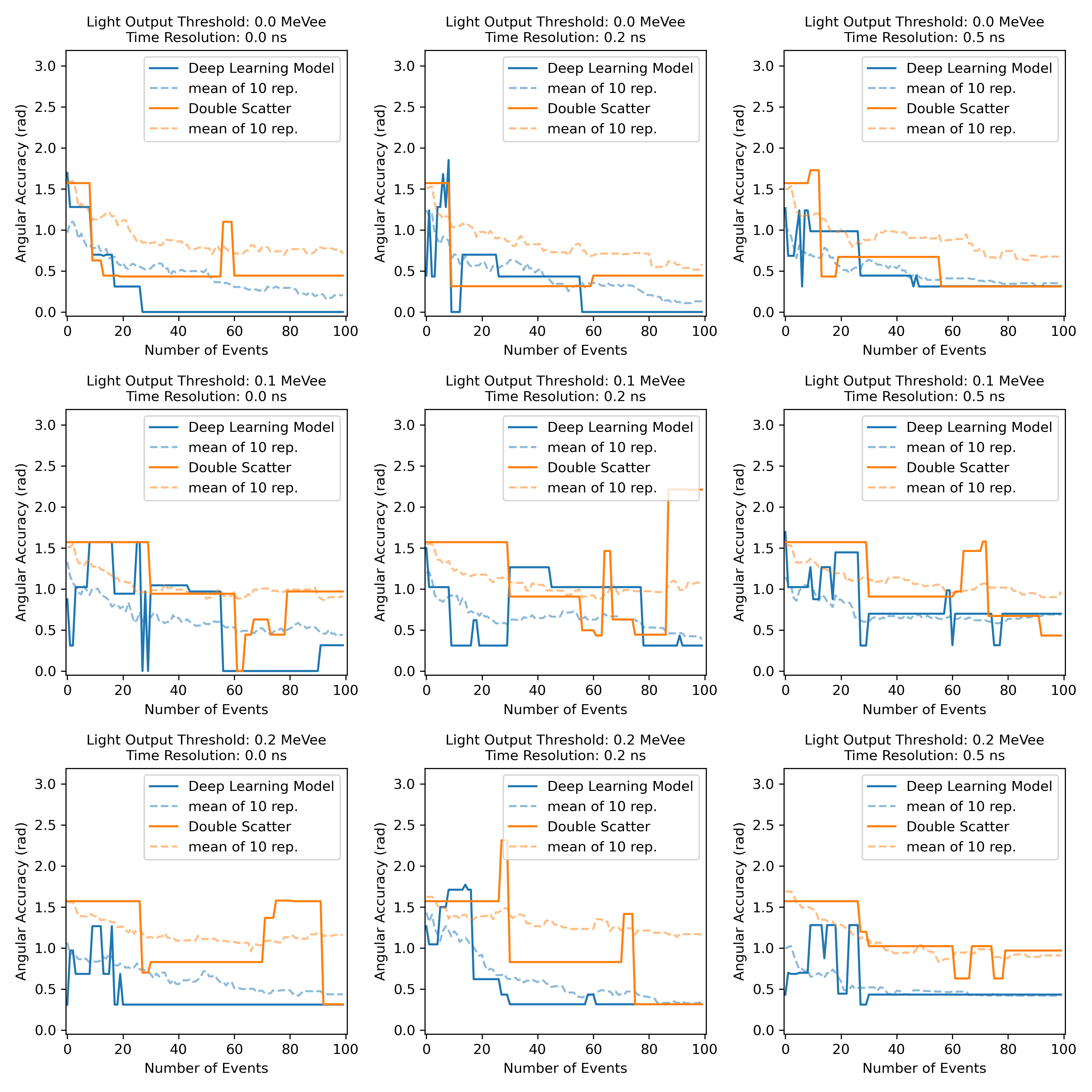}
    \caption{Evolution of angular accuracy using the deep learning model and the double-scatter backprojection method for several different time resolutions. All reconstructions use the same event dataset.}
    \label{fig:9}
\end{figure}

Figure~\ref{fig:9} shows the effect on the time resolution and the light output threshold on the evolution of angular accuracy with event statistics. The true source is $0 < \cos{\theta} < 0.2$ and  $0.628 < \phi < 0.942$~rad which is same direction used in Figure~\ref{fig:7}. The solid lines represent a set of 100 events; the same events are used for all depicted cases. The dashed lines represent a mean of 10 repetitions; 100 events are randomly sampled for each repetition. In the case of the deep learning model, 9 independent models are trained using the data that is processed for each condition. Therefore, the evolution of angular accuracy seems to vary a lot depending on the conditions. In case of the double-scatter backprojection method, the time resolution affects the $\tau$, which affects the estimation of $\theta_{DS}$. On the other hand, the light output threshold can affect the event, which leads to decrease the number of double scatter events or misclassification of double-scatter event by cutting off signals with small light output. Therefore the time resolution primarily affects the angular accuracy derived from a single event, while the light output threshold affects the evolution of the angular accuracy by reducing the number of events from which the information can be extracted. According to the average trend, in the double-scatter backprojection method, when a higher light output threshold and worse time resolution are applied, the angular accuracy is worse. However, in case of deep learning model, this is not always the case, and there are instances where the reconstruction is better even with high light output threshold and time resolution. This may be explained by the fact that deep learning models are trained by the datasets to which these specific conditions apply. 

\section{Conclusion}

In this paper, we investigated the performance of an RNN-based deep learning method in reconstructing the direction of fast neutrons incident onto a segmented scintillation detector. The performance was studied using event data generated through GEANT4 simulation. Data included a series of recoil proton signals generated by an incident fast neutron source, and they were stored as a sequence. The light output thresholds and time resolution was applied to make the data more realistic. To train the model and perform reconstruction from the sequential data,a RNN-based neural network architecture was designed as a deep learning model. A double-scatter method for estimating direction of fast neutron was utilized and compared with the proposed deep learning-based method as a benchmark. The deep learning-based method shows a better angular accuracy, or equivalently requires fewer events to reach the set accuracy. While the single and multiple scatters are discarded in the double-scatter backprojection method, they can provide useful information in case of the deep learning-based method. It is expected that the model can reconstruct the direction of the source even if some of signals are removed by the neutron-gamma discrimination algorithm or by a pulse-height event threshold. This study is still limited in that it uses only simulated data; significantly more insight will be provided when the model is trained and evaluated with real experimental data in the future. 

\section{Acknowledgments}
This work was supported by the Department of Energy National Nuclear Security Administration, Consortium for Monitoring, Verification and Technology (DE-NE000863), Nuclear Global Fellowship Program through the Korea Nuclear International Cooperation Foundation(KONICOF) funded by the Ministry of Science and ICT, and partially supported by the Department of Energy, Nuclear Energy University Program Fellowship.

\bibliography{refs}

\begin{thebibliography}{10}
\expandafter\ifx\csname url\endcsname\relax
  \def\url#1{\texttt{#1}}\fi
\expandafter\ifx\csname urlprefix\endcsname\relax\def\urlprefix{URL }\fi
\expandafter\ifx\csname href\endcsname\relax
  \def\href#1#2{#2} \def\path#1{#1}\fi

\bibitem{direc_vanier2007}
P.~E. Vanier, L.~Forman, I.~Dioszegi, C.~Salwen, V.~J. Ghosh, Calibration and
  testing of a large-area fast-neutron directional detector, in: 2007 IEEE
  Nuclear Science Symposium Conference Record, Vol.~1, IEEE, 2007, pp.
  179--184.

\bibitem{scattercamera_Mascarenhas2009}
N.~Mascarenhas, J.~Brennan, K.~Krenz, P.~Marleau, S.~Mrowka, Results with the
  neutron scatter camera, IEEE Transactions on Nuclear Science 56~(3) (2009)
  1269--1273.
\newblock \href {https://doi.org/10.1109/TNS.2009.2016659}
  {\path{doi:10.1109/TNS.2009.2016659}}.

\bibitem{doublescatter_Steinberger2020}
W.~M. Steinberger, M.~L. Ruch, N.~Giha, A.~D. Fulvio, P.~Marleau, S.~D. Clarke,
  S.~A. Pozzi, \href{https://doi.org/10.1038/s41598-020-58857-z}{Imaging
  special nuclear material using a handheld dual particle imager}, Scientific
  Reports 10~(1) (2020) 1855.
\newblock \href {https://doi.org/10.1038/s41598-020-58857-z}
  {\path{doi:10.1038/s41598-020-58857-z}}.
\newline\urlprefix\url{https://doi.org/10.1038/s41598-020-58857-z}

\bibitem{doubleScatter_braverman2018}
J.~Braverman, J.~Brennan, E.~Brubaker, B.~Cabrera-Palmer, S.~Czyz, P.~Marleau,
  J.~Mattingly, A.~Nowack, J.~Steele, M.~Sweany, et~al., {Single-volume neutron
  scatter camera for high-efficiency neutron imaging and spectroscopy}, arXiv
  preprint arXiv:1802.05261 (2018).

\bibitem{singlevolume_Manfredi2020}
J.~J. Manfredi, E.~Adamek, J.~A. Brown, E.~Brubaker, B.~Cabrera-Palmer,
  J.~Cates, R.~Dorrill, A.~Druetzler, J.~Elam, P.~L. Feng, M.~Folsom,
  A.~Galindo-Tellez, B.~L. Goldblum, P.~Hausladen, N.~Kaneshige, K.~Keefe,
  T.~A. Laplace, J.~G. Learned, A.~Mane, P.~Marleau, J.~Mattingly, M.~Mishra,
  A.~Moustafa, J.~Nattress, K.~Nishimura, J.~Steele, M.~Sweany, K.~Weinfurther,
  K.-P. Ziock, \href{https://doi.org/10.1117/12.2569995}{{The single-volume
  scatter camera}}, in: A.~Burger, S.~A. Payne, M.~Fiederle (Eds.), Hard X-Ray,
  Gamma-Ray, and Neutron Detector Physics XXII, Vol. 11494, International
  Society for Optics and Photonics, SPIE, 2020, p. 114940V.
\newblock \href {https://doi.org/10.1117/12.2569995}
  {\path{doi:10.1117/12.2569995}}.
\newline\urlprefix\url{https://doi.org/10.1117/12.2569995}

\bibitem{doubleScatter_Galindo_Tellez_2021}
A.~Galindo-Tellez, K.~Keefe, E.~Adamek, E.~Brubaker, B.~Crow, R.~Dorrill,
  A.~Druetzler, C.~Felix, N.~Kaneshige, J.~Learned, J.~Manfredi, K.~Nishimura,
  B.~P. Souza, D.~Schoen, M.~Sweany,
  \href{https://doi.org/10.1088/1748-0221/16/04/p04013}{Design and calibration
  of an optically segmented single volume scatter camera for neutron imaging},
  Journal of Instrumentation 16~(04) (2021) P04013.
\newblock \href {https://doi.org/10.1088/1748-0221/16/04/p04013}
  {\path{doi:10.1088/1748-0221/16/04/p04013}}.
\newline\urlprefix\url{https://doi.org/10.1088/1748-0221/16/04/p04013}

\bibitem{direc_BOWDEN20101}
N.~Bowden, M.~Heffner, G.~Carosi, D.~Carter, P.~O’Malley, J.~Mintz, M.~Foxe,
  I.~Jovanovic,
  \href{https://www.sciencedirect.com/science/article/pii/S0168900210019728}{Directional
  fast neutron detection using a time projection chamber}, Nuclear Instruments
  and Methods in Physics Research Section A: Accelerators, Spectrometers,
  Detectors and Associated Equipment 624~(1) (2010) 153--161.
\newblock \href {https://doi.org/https://doi.org/10.1016/j.nima.2010.09.010}
  {\path{doi:https://doi.org/10.1016/j.nima.2010.09.010}}.
\newline\urlprefix\url{https://www.sciencedirect.com/science/article/pii/S0168900210019728}

\bibitem{direc_Igor2009}
I.~Jovanovic, M.~Heffner, L.~Rosenberg, N.~S. Bowden, A.~Bernstein, D.~Carter,
  M.~Foxe, M.~Hotz, M.~Howe, A.~Myers, C.~Winant, Directional neutron detection
  using a time projection chamber, IEEE Transactions on Nuclear Science 56~(3)
  (2009) 1218--1223.
\newblock \href {https://doi.org/10.1109/TNS.2009.2017194}
  {\path{doi:10.1109/TNS.2009.2017194}}.

\bibitem{scatter_STEVANATO2011}
L.~Stevanato, D.~Fabris, X.~Hao, M.~Lunardon, S.~Moretto, G.~Nebbia,
  S.~Pesente, L.~Sajo-Bohus, G.~Viesti,
  \href{https://www.sciencedirect.com/science/article/pii/S0969804310004240}{Light
  output of ej228 scintillation neutron detectors}, Applied Radiation and
  Isotopes 69~(2) (2011) 369--372.
\newblock \href
  {https://doi.org/https://doi.org/10.1016/j.apradiso.2010.10.022}
  {\path{doi:https://doi.org/10.1016/j.apradiso.2010.10.022}}.
\newline\urlprefix\url{https://www.sciencedirect.com/science/article/pii/S0969804310004240}

\bibitem{scint_KASCHUCK2002}
Y.~Kaschuck, B.~Esposito, L.~Trykov, V.~Semenov,
  \href{https://www.sciencedirect.com/science/article/pii/S0168900201014991}{Fast
  neutron spectrometry with organic scintillators applied to magnetic fusion
  experiments}, Nuclear Instruments and Methods in Physics Research Section A:
  Accelerators, Spectrometers, Detectors and Associated Equipment 476~(1)
  (2002) 511--515, int. Workshop on Neutron Field Spectrometry in Science,
  Technolog y and Radiation Protection.
\newblock \href {https://doi.org/https://doi.org/10.1016/S0168-9002(01)01499-1}
  {\path{doi:https://doi.org/10.1016/S0168-9002(01)01499-1}}.
\newline\urlprefix\url{https://www.sciencedirect.com/science/article/pii/S0168900201014991}

\bibitem{doubleScatter_Zhang2016}
X.~Zhang, M.~Zhang, L.~Sheng, Z.~Zhang, K.~Li, B.~Peng, X.~Zhang, X.~Ouyang,
  J.~Liu, J.~Liu, L.~Chen, J.~Zhu, C.~He,
  \href{https://doi.org/10.1007/s11431-015-5922-0}{Image reconstruction of a
  neutron scatter camera}, Science China Technological Sciences 59~(1) (2016)
  149--155.
\newblock \href {https://doi.org/10.1007/s11431-015-5922-0}
  {\path{doi:10.1007/s11431-015-5922-0}}.
\newline\urlprefix\url{https://doi.org/10.1007/s11431-015-5922-0}

\bibitem{machinelearning_2019}
G.~Carleo, I.~Cirac, K.~Cranmer, L.~Daudet, M.~Schuld, N.~Tishby,
  L.~Vogt-Maranto, L.~Zdeborov\'a,
  \href{https://link.aps.org/doi/10.1103/RevModPhys.91.045002}{Machine learning
  and the physical sciences}, Rev. Mod. Phys. 91 (2019) 045002.
\newblock \href {https://doi.org/10.1103/RevModPhys.91.045002}
  {\path{doi:10.1103/RevModPhys.91.045002}}.
\newline\urlprefix\url{https://link.aps.org/doi/10.1103/RevModPhys.91.045002}

\bibitem{machinelearning_Schwartz2021}
M.~D. Schwartz, \href{https://doi.org/10.1162\%2F99608f92.beeb1183}{Modern
  machine learning and particle physics}, Harvard Data Science Review (mar
  2021).
\newblock \href {https://doi.org/10.1162/99608f92.beeb1183}
  {\path{doi:10.1162/99608f92.beeb1183}}.
\newline\urlprefix\url{https://doi.org/10.1162\%2F99608f92.beeb1183}

\bibitem{deeplearning_Shengdong2018}
S.~Du, T.~Li, S.-J. Horng, Time series forecasting using sequence-to-sequence
  deep learning framework, in: 2018 9th International Symposium on Parallel
  Architectures, Algorithms and Programming (PAAP), 2018, pp. 171--176.
\newblock \href {https://doi.org/10.1109/PAAP.2018.00037}
  {\path{doi:10.1109/PAAP.2018.00037}}.

\bibitem{deeplearning_GOMEZFERNANDEZ2021}
M.~Gomez-Fernandez, W.-K. Wong, A.~Tokuhiro, K.~Welter, A.~M. Alhawsawi,
  H.~Yang, K.~Higley,
  \href{https://www.sciencedirect.com/science/article/pii/S016890022031322X}{Isotope
  identification using deep learning: An explanation}, Nuclear Instruments and
  Methods in Physics Research Section A: Accelerators, Spectrometers, Detectors
  and Associated Equipment 988 (2021) 164925.
\newblock \href {https://doi.org/https://doi.org/10.1016/j.nima.2020.164925}
  {\path{doi:https://doi.org/10.1016/j.nima.2020.164925}}.
\newline\urlprefix\url{https://www.sciencedirect.com/science/article/pii/S016890022031322X}

\bibitem{deeplearning_carminati2017}
F.~Carminati, G.~Khattak, M.~Pierini, S.~Vallecorsafa, A.~Farbin, B.~Hooberman,
  W.~Wei, M.~Zhang, B.~Pacela, M.~S. Vitorial, et~al., Calorimetry with deep
  learning: particle classification, energy regression, and simulation for
  high-energy physics, in: Workshop on deep learning for physical sciences
  (DLPS 2017), NIPS, 2017, p.~31.

\bibitem{knoll2010radiation}
G.~F. Knoll, Radiation detection and measurement, John Wiley \& Sons, 2010.

\bibitem{Geant4_AGOSTINELLI2003}
S.~Agostinelli, J.~Allison, K.~Amako, J.~Apostolakis, H.~Araujo, P.~Arce,
  M.~Asai, D.~Axen, S.~Banerjee, G.~Barrand, F.~Behner, L.~Bellagamba,
  J.~Boudreau, L.~Broglia, A.~Brunengo, H.~Burkhardt, S.~Chauvie, J.~Chuma,
  R.~Chytracek, G.~Cooperman, G.~Cosmo, P.~Degtyarenko, A.~Dell'Acqua,
  G.~Depaola, D.~Dietrich, R.~Enami, A.~Feliciello, C.~Ferguson, H.~Fesefeldt,
  G.~Folger, F.~Foppiano, A.~Forti, S.~Garelli, S.~Giani, R.~Giannitrapani,
  D.~Gibin, J.~{Gómez Cadenas}, I.~González, G.~{Gracia Abril}, G.~Greeniaus,
  W.~Greiner, V.~Grichine, A.~Grossheim, S.~Guatelli, P.~Gumplinger,
  R.~Hamatsu, K.~Hashimoto, H.~Hasui, A.~Heikkinen, A.~Howard, V.~Ivanchenko,
  A.~Johnson, F.~Jones, J.~Kallenbach, N.~Kanaya, M.~Kawabata, Y.~Kawabata,
  M.~Kawaguti, S.~Kelner, P.~Kent, A.~Kimura, T.~Kodama, R.~Kokoulin,
  M.~Kossov, H.~Kurashige, E.~Lamanna, T.~Lampén, V.~Lara, V.~Lefebure,
  F.~Lei, M.~Liendl, W.~Lockman, F.~Longo, S.~Magni, M.~Maire, E.~Medernach,
  K.~Minamimoto, P.~{Mora de Freitas}, Y.~Morita, K.~Murakami, M.~Nagamatu,
  R.~Nartallo, P.~Nieminen, T.~Nishimura, K.~Ohtsubo, M.~Okamura, S.~O'Neale,
  Y.~Oohata, K.~Paech, J.~Perl, A.~Pfeiffer, M.~Pia, F.~Ranjard, A.~Rybin,
  S.~Sadilov, E.~{Di Salvo}, G.~Santin, T.~Sasaki, N.~Savvas, Y.~Sawada,
  S.~Scherer, S.~Sei, V.~Sirotenko, D.~Smith, N.~Starkov, H.~Stoecker,
  J.~Sulkimo, M.~Takahata, S.~Tanaka, E.~Tcherniaev, E.~{Safai Tehrani},
  M.~Tropeano, P.~Truscott, H.~Uno, L.~Urban, P.~Urban, M.~Verderi, A.~Walkden,
  W.~Wander, H.~Weber, J.~Wellisch, T.~Wenaus, D.~Williams, D.~Wright,
  T.~Yamada, H.~Yoshida, D.~Zschiesche,
  \href{https://www.sciencedirect.com/science/article/pii/S0168900203013688}{Geant4—a
  simulation toolkit}, Nuclear Instruments and Methods in Physics Research
  Section A: Accelerators, Spectrometers, Detectors and Associated Equipment
  506~(3) (2003) 250--303.
\newblock \href {https://doi.org/https://doi.org/10.1016/S0168-9002(03)01368-8}
  {\path{doi:https://doi.org/10.1016/S0168-9002(03)01368-8}}.
\newline\urlprefix\url{https://www.sciencedirect.com/science/article/pii/S0168900203013688}

\bibitem{plastics_SC_PVT}
D.~Groom, Atomic and nuclear properties of polyvinyltoluene,
  \url{https://pdg.lbl.gov/2022/AtomicNuclearProperties/HTML/polyvinyltoluene.html},
  (Accessed: 11.23.2022) (2021).

\bibitem{sandd_sutanto2021}
F.~Sutanto, T.~Classen, S.~Dazeley, M.~Duvall, I.~Jovanovic, V.~Li, A.~Mabe,
  E.~Reedy, T.~Wu,
  \href{https://www.sciencedirect.com/science/article/pii/S0168900221003934}{Sandd:
  A directional antineutrino detector with segmented 6li-doped
  pulse-shape-sensitive plastic scintillator}, Nuclear Instruments and Methods
  in Physics Research Section A: Accelerators, Spectrometers, Detectors and
  Associated Equipment 1006 (2021) 165409.
\newblock \href {https://doi.org/https://doi.org/10.1016/j.nima.2021.165409}
  {\path{doi:https://doi.org/10.1016/j.nima.2021.165409}}.
\newline\urlprefix\url{https://www.sciencedirect.com/science/article/pii/S0168900221003934}

\bibitem{Direc_Fu2020}
Y.~Fu, Y.~Tian, Y.~Li, J.~Yang, J.~Li,
  \href{https://www.sciencedirect.com/science/article/pii/S0168900218314372}{Directional
  fast neutron detection using a time projection chamber and plastic
  scintillation detectors}, Nuclear Instruments and Methods in Physics Research
  Section A: Accelerators, Spectrometers, Detectors and Associated Equipment
  954 (2020) 161445, symposium on Radiation Measurements and Applications XVII.
\newblock \href {https://doi.org/https://doi.org/10.1016/j.nima.2018.10.123}
  {\path{doi:https://doi.org/10.1016/j.nima.2018.10.123}}.
\newline\urlprefix\url{https://www.sciencedirect.com/science/article/pii/S0168900218314372}

\bibitem{angularacc_miniTimeCube2019}
G.~R. Jocher, J.~Koblanski, V.~A. Li, S.~Negrashov, R.~C. Dorrill,
  K.~Nishimura, M.~Sakai, J.~G. Learned, S.~Usman,
  \href{https://doi.org/10.1063/1.5079429}{minitimecube as a neutron scatter
  camera}, AIP Advances 9~(3) (2019) 035301.
\newblock \href {http://arxiv.org/abs/https://doi.org/10.1063/1.5079429}
  {\path{arXiv:https://doi.org/10.1063/1.5079429}}, \href
  {https://doi.org/10.1063/1.5079429} {\path{doi:10.1063/1.5079429}}.
\newline\urlprefix\url{https://doi.org/10.1063/1.5079429}

\bibitem{sandd_Li2022}
V.~A. Li, F.~Sutanto, T.~M. Classen, S.~A. Dazeley, I.~Jovanovic, T.~C. Wu,
  Evaluation of a positron-emission-tomography-based sipm readout for compact
  segmented neutron imagers, arXiv preprint arXiv:2202.07759 (2022).

\bibitem{RNN_Manaswi2018}
N.~K. Manaswi, \href{https://doi.org/10.1007/978-1-4842-3516-4$\_$9}{RNN and
  LSTM}, Apress, Berkeley, CA, 2018, Ch.~9, pp. 115--126.
\newblock \href {https://doi.org/10.1007/978-1-4842-3516-4$\_$9}
  {\path{doi:10.1007/978-1-4842-3516-4$\_$9}}.
\newline\urlprefix\url{https://doi.org/10.1007/978-1-4842-3516-4$\_$9}

\end{thebibliography}

\end{document}